\documentclass{article}
\usepackage{spconf,amsmath,graphicx,url,times,booktabs,tabularx}
 \usepackage{cite}
\usepackage{comment}
\usepackage{amssymb}
\usepackage{csquotes}
\title{1-D CNN based acoustic scene classification via reducing layer-wise dimensionality}

\name{Arshdeep Singh \sthanks{The author conducted the current research partly during his PhD. The author acknowledges IIT Mandi, India for their computational resources.}}
     
\address{Research Fellow, Department of Electrical and Electronic Engineering,\\ University of Surrey, UK.\\
arshdeep.singh@surrey.ac.uk         
}

\begin{document}

\ninept
\maketitle

\begin{sloppy}

\begin{abstract}
This paper presents an alternate representation framework to commonly used time-frequency representation for acoustic scene classification (ASC). A raw audio signal is represented using a pre-trained convolutional neural network (CNN) using its various intermediate
layers. The study assumes that the representations obtained from the intermediate layers lie in low-dimensions intrinsically. To obtain
low-dimensional embeddings, principal component analysis is performed, and the study analyzes that only a few principal components are significant. However, the appropriate number of significant
components are not known. To address this, an automatic dictionary learning framework is utilized that approximates the underlying subspace.  Further, the low-dimensional embeddings are aggregated in a late-fusion manner in the ensemble framework to incorporate hierarchical information learned at various intermediate
layers. The experimental evaluation is performed on publicly available DCASE 2017 and 2018 ASC datasets on a pre-trained 1-D CNN, SoundNet. Empirically, it is observed that deeper layers show more compression ratio\footnote{A compression ratio is defined as a ratio of reduced dimensionality and the actual dimensionality.} than others. At 70\% compression ratio across different datasets, the performance is similar to that obtained
without performing any dimensionality reduction. The
proposed framework outperforms the time-frequency representation based methods.

\end{abstract}


\section{Introduction}
\label{sec:intro}
Acoustic scene classification (ASC) utilizes sounds produced in the physical environment to classify the underlying scene \cite{barchiesi2015acoustic}. Initially, the hand-engineered features such as Mel-frequency cepstral coefficients (MFCCs) and log-mel energies inspired from speech and music recognition fields have been utilized for ASC. As hand-engineered features are not adaptive and do not generalize well, recent techniques employ learning-based approaches such as deep learning \cite{han2017convolutional, hershey2017cnn, arandjelovic2017look, aytar2016soundnet} and dictionary learning  \cite{rakotomamonjy2017supervised, singh2018ape,  kong2012dictionary}. Typically, the learning-based frameworks use time-frequency representations of an audio signal to learn discriminative representations. The time-frequency representations involve short-time Fourier transform, which may not be sufficient to represent diverse sound activities such as slowly varying sounds (sea waves, music) and impulsive sounds (gunshot, clock tik) both at the same time. This is due to the fixed time-frequency resolution of the  Fourier atoms. This study analyzes how well the representations obtained using a learned basis perform compared to commonly used time-frequency representations.

Only a few studies \cite{aytar2016soundnet, Purohit2018,eusipco_paper} apply convolution neural networks (CNNs) to learn representations from the raw audio directly.  The study \cite{eusipco_paper} utilizes representations obtained from various intermediate layers of a pre-trained neural network SoundNet  in the ensemble framework. As large-scale networks are trained  on large-scale datasets, it may be possible that the network is over-parameterized for the underlying problem due to redundancy \cite{denton2014exploiting}. 

\begin{figure}[t]
    \centering
    \includegraphics[scale=0.43]{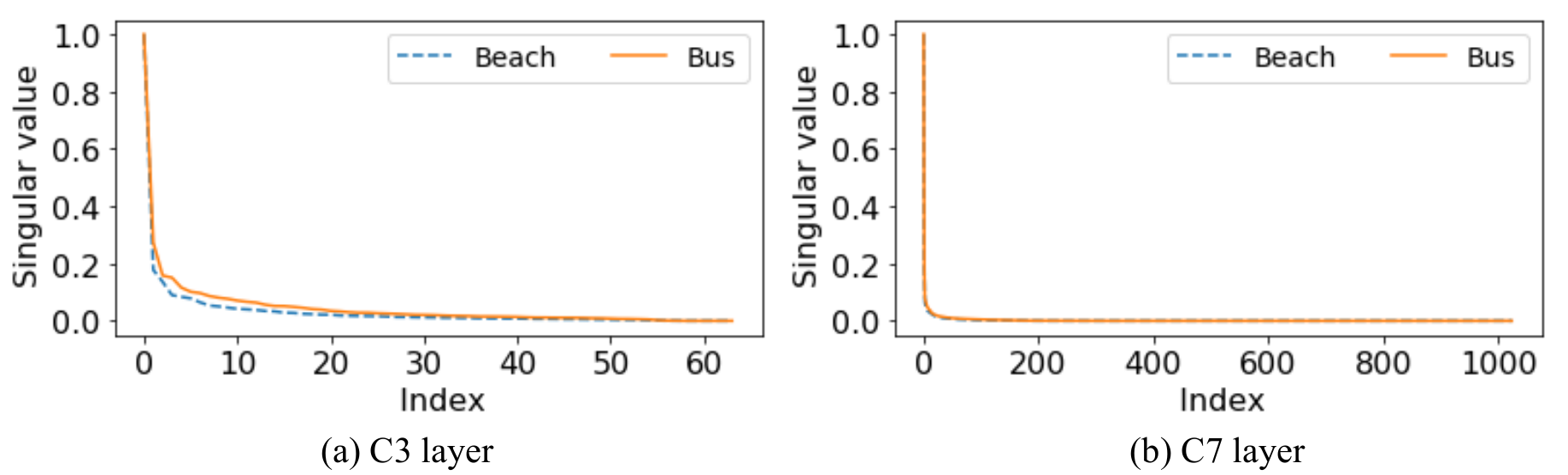}
    \caption{Normalized singular values obtained for beach and bus acoustic scenes when representations are obtained from  third convolution layer (C3) having 64 filters and seventh convolution layer (C7) having 1024 filters in SoundNet. The acoustic scene classes are taken from DCASE ASC 2017 development dataset \cite{Heittola2017}}.
    \label{fig: svd_values}
\end{figure}

This study also focuses on reducing dimensionality of intermediate layer representations in the ensemble framework \cite{eusipco_paper}.  For this, the work assumes that the responses produced by the filters in a given intermediate layer have an intrinsic low-dimension structure. This is based on the hypothesis that there is redundancy in the network \cite{frankle2018lottery,livni2014computational} and a subset of filters or parameters may be sufficient to represent audio signals for the underlying problem as a large-scale dataset is used to train SoundNet. Empirically, it can be seen in Figure \ref{fig: svd_values} which shows that the representations in the intermediate layers result into rapid decay in their singular values. This suggests that the embeddings have a low-dimensional structure. In the context of the ensemble framework, reducing layer-wise dimensionality decreases the computational complexity during learning and inference of the layer-wise classifiers. However, the appropriate dimensionality of the embedding space is unknown. 

To identify the dimensionality of the intermediate layers, a dictionary learning framework is employed, which selects the appropriate dimensionality of the embedding space by minimizing the reconstruction error and the correlation among dictionary atoms during the learning process itself.

The major contributions of this paper are summarized as follows:

\begin{itemize}
    \item An alternate feature representation framework for ASC that uses intermediate layers of SoundNet to commonly used time-frequency representations.
   \item To select the appropriate dimensionality of the embeddings in SoundNet, a dictionary learning framework is used.
   \item Our experiments reveal that the deeper layers show more compression ratio than other layers.
\end{itemize}

The rest of this paper is organized as follows. First, a brief overview of the ensemble framework using SoundNet is presented in Section \ref{sec: breif overview on SoundNet}. Second, the proposed framework to identify layer-wise dimensionality is described in Section \ref{sec: proposed methodology}. Next, Section  \ref{sec: performance eval} includes performance evaluation. Finally, the conclusion is presented in Section \ref{sec: conclusion}.

\section{A brief overview about ensemble framework on SoundNet}
\label{sec: breif overview on SoundNet}

SoundNet \cite{aytar2016soundnet} is a pre-trained one dimensional (1-D)\footnote{Feature maps and filters are 1-D.} CNN, which is trained on a raw audio signal using a teacher-student based transfer learning framework. The architecture of SoundNet is shown in Figure \ref{fig: soundnet}. Different intermediate layers of SoundNet represent different sound-related concepts. For example, the middle layer (fifth convolution layer) represents concepts such as music-tone like, babycry like, tapping like, etc. The seventh convolution layer represents  high-level concepts such as music like, voice like, etc. The convolution layer consists of \enquote{p-conv} and\enquote{conv} layers. Here, conv represents convolutional layer with ReLU activation, and  p-conv layer represents the convolution layer without ReLU activation.

In \cite{eusipco_paper}, an ensemble framework is proposed to incorporate various features learned at different intermediate layers of SoundNet. The feature maps from a given intermediate layer are transformed into aggregated representations (fixed-length feature vector) by applying a global sum pooling operation. An  illustration to obtain aggregated representations from $l^{th}$ intermediate layer  is shown in Figure \ref{fig: layer-wise embeddings}. Next, a classifier is learned for each intermediate layer independently. Finally, the probability scores from each classifier are aggregated to compute the ultimate classification scores.

This paper utilizes the ensemble framework \cite{eusipco_paper} and aims to reduce the dimensionality of intermediate layer features without degrading the performance.

\begin{figure}[t]
\includegraphics[scale=0.46]{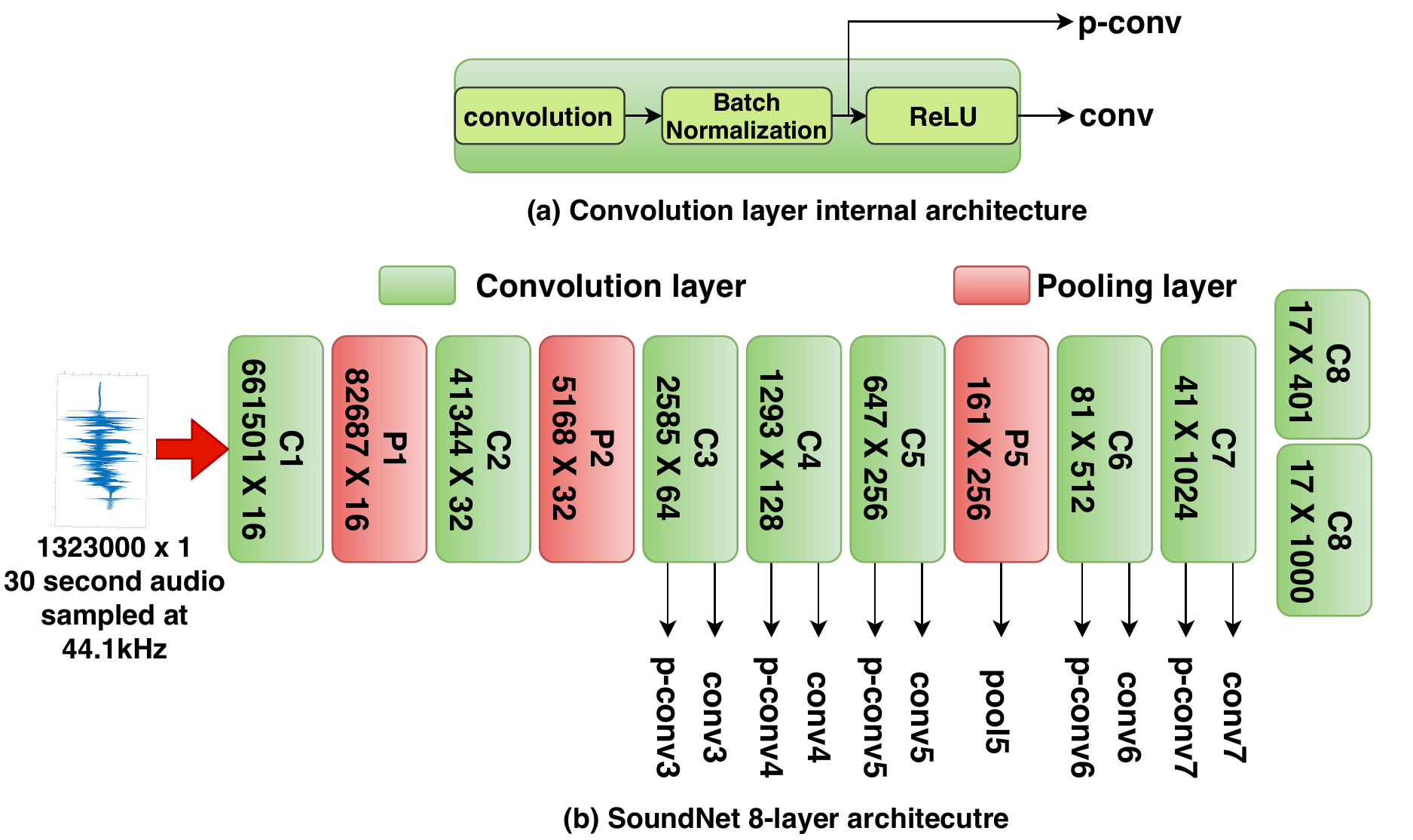}
\caption{SoundNet architecture \cite{aytar2016soundnet}. (a) convolution layer architecture, \textit{conv} denotes the output of  convolution layer output and \textit{p-conv} denotes the output of batch normalization layer. (b) 8-layer architecture with \textrm{CX} and \textrm{PX}  as $\textrm{X}^{th}$ convolution and pooling layers respectively. }\label{fig: soundnet}
\end{figure}

\section{Proposed methodology}
\label{sec: proposed methodology}

Consider a 1-D  CNN having $L$-layers with indexes $\in$ \{1,2,$\dotsi$,$l$,$\dotsi$,$L$\}\footnote{This method can be extended to other CNNs (2-D) as well.}. Let there be $n_l$ number of feature maps, each of length $s_l$ in the $l^{th}$ layer. Let  $\mathbf{X}_l$ of size  $(n_{l} \times s_{l})$ denotes  the  feature map matrix, produced by stacking all the feature maps  in the  $l^{th}$ layer. Let there are  $\textrm{C}$ acoustic scene classes, and $M_c$ denotes the number of acoustic examples in a $c^{th}$ class. Let $M$ represents a total number of acoustic examples in all classes.  

Let $\mathbf{H}_l \in \mathbb{R}^{n_l \times 1}$ denotes the aggregated intermediate representation obtained in the $l^{th}$ layer by performing a global sum pooling operation across feature maps on $\mathbf{X}_l$. 
%An illustration of obtained aggregated representation is shown in Figure xxx.

\begin{figure}[ht]
    \centering
    \includegraphics[scale=0.26]{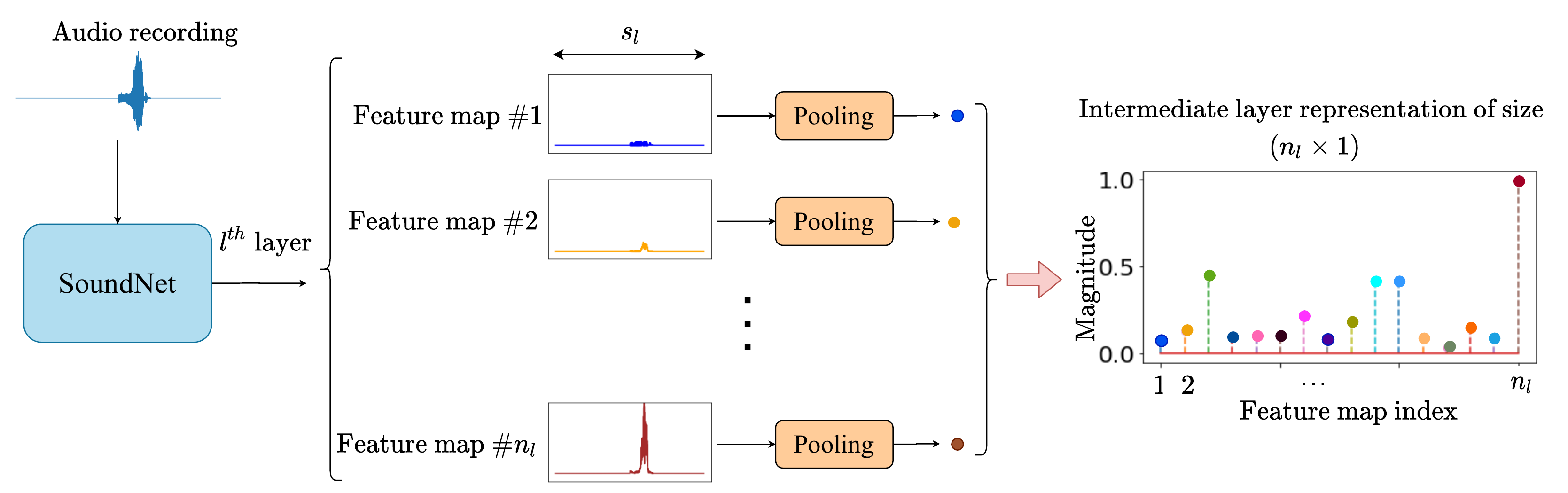}    
    \caption{An illustration of aggregated representation obtained from $l^{th}$ intermediate layer of SoundNet. Here $n_l$ represents number of feature maps each having $s_l$ length.}    \label{fig: layer-wise embeddings}
\end{figure}

Let $\mathbf{Y}_c \in  \mathbb{R}^{n_l \times M_c}$ denotes a $c^{th}$ class-specific data matrix obtained by stacking aggregated intermediate representations of  the $c^{th}$ class examples in the $l^{th}$ layer. Let a data matrix $\mathbf{Y}_l = \mathbf{[Y_{\textrm{1}}, Y_{2}, ...,Y_{\textrm{c}},..,Y_{\textrm{C}}}] \in \mathbb{R}^{n_l \times M}$, is obtained by stacking  the class-specific data matrices of all classes.

%This section explain the dictionary learning frameworks to compute CCDE for an acoustic recording. Subsequently, intra-ensemble framework is described. 

\subsection{Obtaining low-dimensional embeddings} Given $\mathbf{Y}_l \in \mathbb{R}^{n_l \times M}$, our aim is to compute embeddings $\mathbf{Z}_l = \mathbf{T_l} \mathbf{Y}_l, \in \mathbb{R}^{d_l \times M}$ using a transformation $\mathbf{T_l} \in \mathbb{R}^{d_l \times n_l}$ with $ d_l << n_l$ for $l^{th}$ layer.

To obtained $\mathbf{T}_l : \mathbb{R}^{n_l} \longrightarrow \mathbb{R}^{d_l}$, principal component analysis (PCA) is performed on  $\mathbf{Y}_l$. Here, $d_l$ is a hyper-parameter and commonly it is chosen as the number of significant singular values that represents a given variance of the underlying data. However, it may result into overestimation of dimensionality as the discriminative information is not accounted for computing low-dimensional representations. This study proposes to utilize a dictionary learning framework to identify $d_l$.

%In this subsection, we explain the approaches to obtain CCE for $l^{th}$ layer. For this, first, a given acoustic recording is represented into $l^{th}$ layer, aggregated intermediate representations, $H_l$ and then, a linear transformation is performed on  $H_l$ to obtain compact representations as explained below. 
%\subsubsection{Pair-wise distance preserving transformation}
\label{sec: random sketching algorithm}

%The pairwise distance between a set of data points can be preserved in the low-dimensional embedding space with high probability using a random linear transform \cite{larsen2014johnson,fedoruk2018dimensionality}. The transformation, $\mathbf{T}_l : \mathbb{R}^{n_l} \longrightarrow \mathbb{R}^{d_l}$, is obtained by utilizing a random sketching algorithm \cite{li2018randomly} (we refer it as RSA). The RSA aims to choose a union of features ($d_l \subset n_l$) randomly from the data matrix $\mathbf{Y}_l$.

%by preserving the pairwise-distance in the embedding space with high probability [cite:J-L lemma].
%The transformation matrix $\mathbf{T}_l$ is obtained  using  Equation \ref{Equ: T=QP}. 

\begin{figure*}[t]
\centering
\includegraphics[scale=0.36]{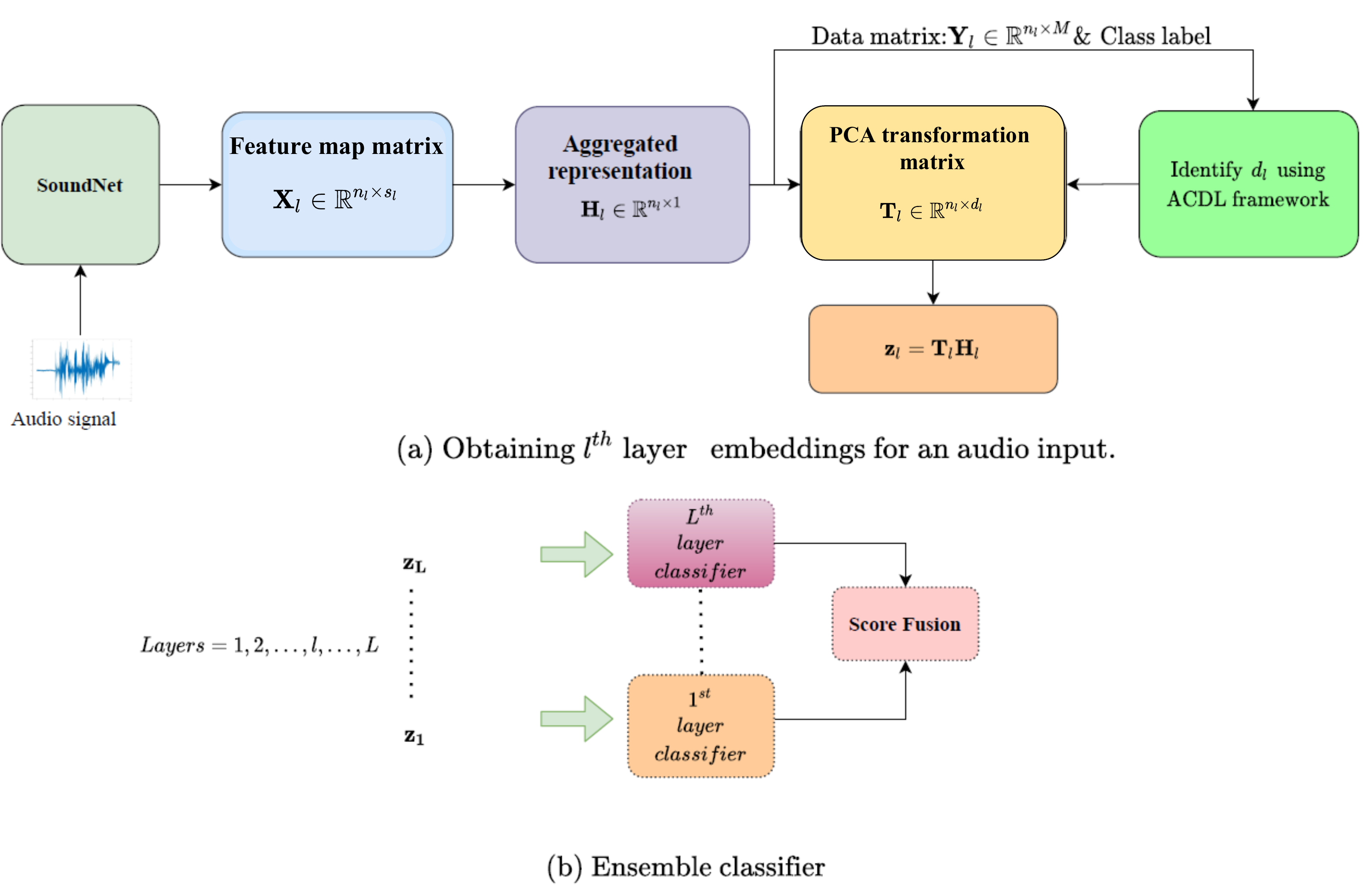}
%\vspace{-0.5cm}
\caption{An overall proposed framework is shown. (a) The procedure to obtain embeddings for an acoustic signal from $l^{th}$ layer of SoundNet. (b) The ensemble framework is shown which utilizes embeddings from various intermediate layers. Here $z_l$ denotes the one of the columns of $\mathbf{Z}_l$.}
\label{fig: Overall proposed}
\end{figure*}

\subsection{Obtaining layer-wise optimal dimensions ($d_l$) using ACDL} 

An automatic compact dictionary learning (ACDL) \cite{song2016dictionary} method is employed to obtain the dimensionality of the embedding space in the $l^{th}$ layer. The ACDL-method gives a compact dictionary by optimizing the objective function as given in Equation \ref{Equ: ACDL}. The optimization process jointly minimizes reconstruction error,  classification loss and eliminates redundant atoms in the dictionary. The number of optimal dictionary atoms $(\mathbf{A}^*)$ thus obtained is chosen as the dimensionality of the embedding space.

\label{sec: layer-wise optimal ACDL}
\begin{equation}
\label{Equ: ACDL}
\resizebox{1\hsize}{!}{
\begin{tabular}{c}
$[ \mathbf{A}^*, \mathbf{W}^* ]=\arg\underset{\mathbf{A},\mathbf{W}}{\min } \Vert\mathbf{Y}-\mathbf{A}\mathcal{F}(\mathbf{A},\mathbf{W})\mathbf{Z}\Vert_F^2+ \gamma \Vert\mathbf{G}-\mathbf{W}\mathcal{F}(\mathbf{A},\mathbf{W})\mathbf{Z}\Vert^2_F$ \\
$+\lambda \Vert \mathbf{Z}\Vert_1 \hspace{0.5cm}  \text{s.t.}  \hspace{0.2cm}\mathbf{Z}\succeq 0,\;\sum \mathbf{z}_i=1 $
\end{tabular}
}
\end{equation}

\begin{equation}
\label{Equ: FDW}
\resizebox{1\hsize}{!}{$%
\begin{aligned} 
  \mathcal{F}(\mathbf{A},\mathbf{W})_{jj} = \left\{ 
    \begin{array}{cl} \mathcal{F}(\mathbf{A},\mathbf{W})_{jj},\;&{}\mathbf{a}_j=0,\mathbf{a}_j\in \mathbf{A_c}\\ \\ 0, \;&{} \displaystyle \frac{\mathop{\text {min}}\limits _{k}||\mathbf{a}_j-\mathbf{a}_k||_2}{\mathop{\text {max}}\limits _{p,q}||\mathbf{a}_p-\mathbf{a}_q||_2}<\tau \\ ~\; &{} \text {and } \mathbf{w}_j\ln \mathbf{w}_j<\mathbf{w}_k\ln \mathbf{w}_k\\ &{} \mathbf{w}_j,\mathbf{w}_k\in \mathbf{W_c}\\ ~\; &{} \mathbf{a}_*\in \mathbf{A_c}, \mathbf{a}_*\ne 0\\ &{} or ~\mathcal{F}(\mathbf{A},\mathbf{W})_{jj} =0 \\ 1, \;&{} \text{otherwise} 
    \end{array}\right.
\end{aligned}
$%
}
\end{equation}

In Equation \ref{Equ: ACDL}, the first term represents reconstruction error, the second term denotes the classification loss, and $\mathbf{Z}$ denotes the non-negative convex coefficients. $\mathbf{G}$ represents class-information of each column in $\mathbf{Y}$ in the form of a one-hot encoding vector, and $\mathbf{A}$ denotes a dictionary and $\mathbf{W}$ is a weight matrix. $\mathbf{A}^*$ and $\mathbf{W}^*$ represents optimal dictionary and the weight matrix respectively obtained after optimization process. $\lambda$ and $\gamma$ are the regularizes. $\tau \in [0,1]$ denotes a threshold to eliminate similar atoms. An alternate-minimization procedure is applied to obtain the optimal dictionary. The redundant elements are eliminated by using an indicator function $\mathcal{F}(\mathbf{A},\mathbf{W})$ as given in Equation \ref{Equ: FDW}. The indicator function removes highly correlated dictionary atoms as measured in Euclidean distance,  which shows less discrimination as measured in entropy. The $\mathbf{A_c}$ and $\mathbf{W_c}$ are equivalent to $\mathbf{A}$ and $\mathbf{W}$ respectively, but the redundant atoms that do not belong to the $c^{th}$ class are set to zero.

\noindent \textbf{Ensemble framework:} After obtaining embeddings from intermediate layers, a classifier is trained for each of the intermediate layers. An ensemble framework combines the probabilities scores from each of the classifiers in a later-fusion manner. The overall proposed framework is shown in Figure \ref{fig: Overall proposed}.

\section{Performance Evaluation}
\label{sec: performance eval}

In this section, we present the dataset used, experimental details and performance analysis using the proposed framework.

\subsection{Datasets used and Experimental setup }

The proposed framework is evaluated on  fine-tuned SoundNet using various acoustic scene classification datasets, which include; %(a) TUT DCASE 2016 ASC  (DCASE-16) development and evaluation dataset \cite{mesaros2016tut}, comprises of 15 acoustic scene classes, 
(a) TUT DCASE Acoustic scenes 2017, development \cite{mesaros2017dcase} (DCASE-17) comprises of 15 acoustic scene classes and 
(b)  TUT Urban Acoustic Scenes 2018 development (DCASE-18) \cite{mesaros2018multi} consists of 10 acoustic scenes. The fine-tuning process is performed end-to-end using C1 to C7 intermediate layers of SoundNet followed by a fully connected layer and a classification layer for both datasets separately.
%, (d) ESC-10, comprises of 10 acoustic scene classes and (e) ESC-50 \cite{piczak2015esc}, comprises of 50 acoustic scene classes.  

After fine-tuning,  $L$ $\in$ \{p-conv3, conv3, $\dotsi$ , conv7\} layers, a total of 10 layers, are utilized in the ensemble framework.  The classification at each intermediate layer is performed using a non-linear SVM with a polynomial kernel.  The hyper-parameters of the SVM are selected by cross-validation. The performance is measured in terms of accuracy.  
The performance is computed over multiple folds as given in each dataset. The baseline\footnote{A baseline is an ensemble framework without any dimensionality reduction.}   accuracy obtained  for  DCASE-17 and DCASE-18  datasets is   76.12\%, and 62.98\%.

In ACDL framework, the reconstruction error equals 0.01 is chosen as the stopping criterion. The deeper layers meet the stopping criterion at $\tau$ $\ge$ 0.7. On the other hand,  $\tau$ varies between 0.3 and 0.5 for shallow layers to achieve the desired stopping criterion.

The effectiveness of selecting layer-wise dimensionality using the proposed method is compared with that of   PCA-based explained variance ratio\footnote{PCA-based explained variance ratio is the ratio of  a number principal components that represents the given variance of the underlying data points to the total number of components.} based dimensionality selection. %that  represents number of principal components that explain the variance of the underlying data set at the given ratio. 
The ratio is varied from 85\% to 100\%. A relatively low  explained variance ratio yields more compression ratio.

\subsection{Results and analysis}

\begin{figure}[ht]
    \centering
    \includegraphics[scale=0.5]{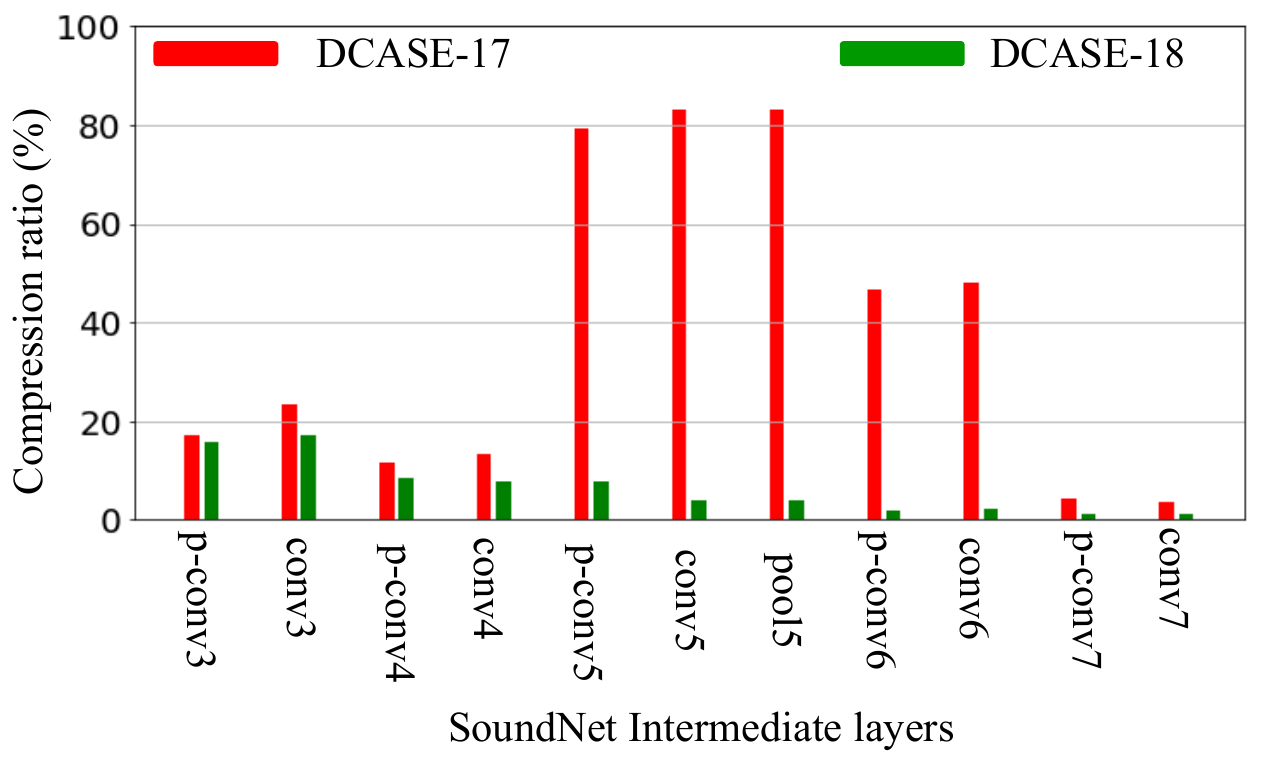}
 %   \vspace{-0.5cm}
   \caption{Layer-wise compression ratio obtained in SoundNet for DCASE-17 and DCASE-18 datasets using the proposed method.}
   \label{fig: compression_ratio}
\end{figure} 

 \begin{figure}[ht]
     \centering
     \includegraphics[scale=0.5]{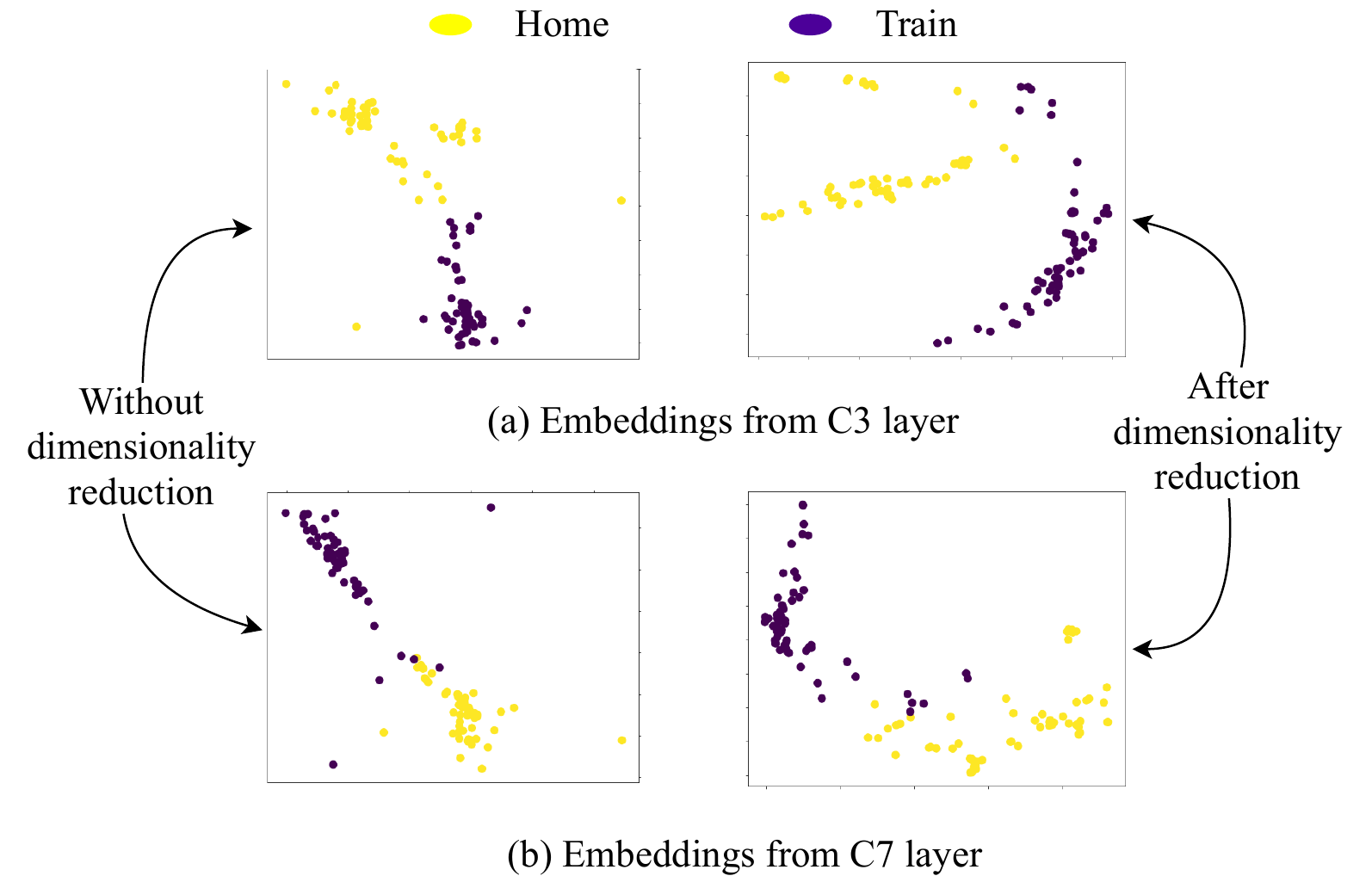}

     \caption{t-SNE plot of embeddings obtained from SoundNet before and after dimensionality reduction for \enquote{home} and \enquote{train} scenes taken from DCASE-17 dataset.}
     \label{fig: tSNE}
 \end{figure}

 \begin{figure}[ht]
     \includegraphics[scale=0.42]{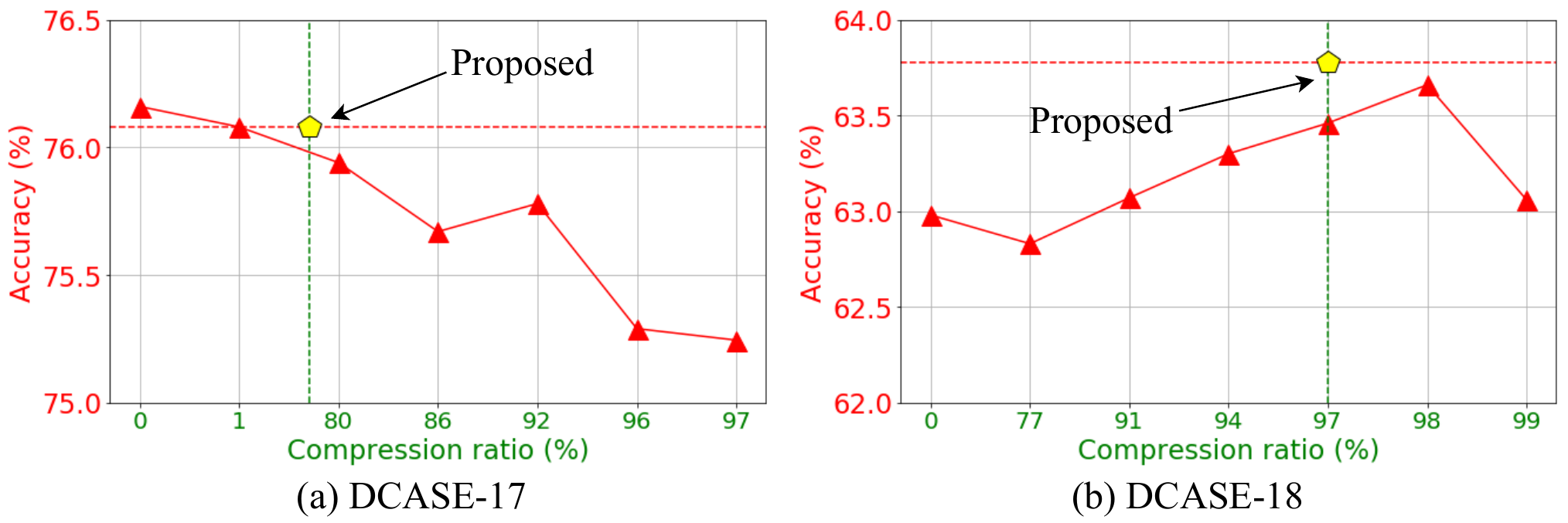}
     \caption{Accuracy and compression ratio obtained for DCASE-17 and DCASE-18 is shown. Different compression is obtained by varying the PCA-based explained variance ratio( $\in\{100\%,99.9\%,99\%,98\%,95\%,90\%,85\%\}$). Here 100\% ratio means no compression or 0\% compression ratio, and 85\% ratio results into 97\%,99\% compression ratio for DCASE-17 and DCASE-18 respectively. The proposed method selects compression ratio using ACDL based dictionary learning framework.}
     \label{fig: acc_compresion}
 \end{figure}

The compression ratio obtained using the proposed method across various intermediate layers of SoundNet is shown in Figure \ref{fig: compression_ratio}. The deeper layers (p-conv7, conv7) show a high compression ratio ($\ge$ 95\%) compared to other layers. This suggests that the aggregated representations from the deeper layer are more sparse than the shallow layers.

 \begin{figure}[ht]
    \centering
    \includegraphics[scale=0.4]{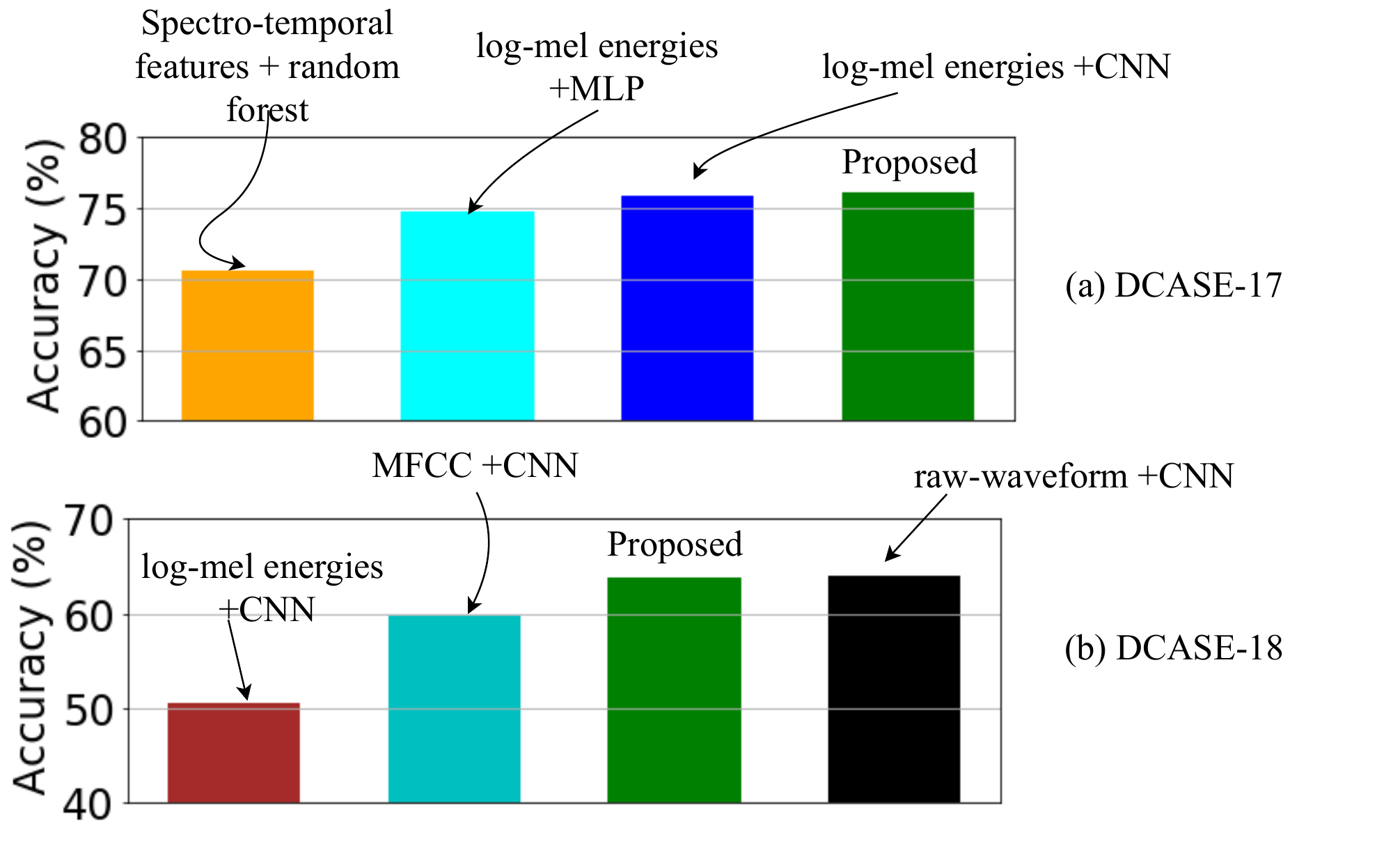}
  %  \vspace{-0.5cm}
    \caption{Comparison of accuracy for DCASE-17 and DCASE-18 dataset with similar methods that utilizes time-frequency representations.}
    \label{fig: acc_comparison}
\end{figure}

Figure \ref{fig: tSNE} shows the embeddings obtained before and after  reducing layer-wise dimensionality using the proposed method from C3 and C7 convolutional layer. It can be observed that the discrimination between different scene classes is preserved in the low-dimensional embedding space.

The accuracy obtained at different compression ratio\footnote{Different compression ratio is obtained by varying PCA-based explained variance ratio.} for DCASE-17 and DCASE-18 dataset is shown in Figure \ref{fig: acc_compresion}. Also, the accuracy and the compression ratio obtained using the proposed method is shown. The proposed method gives 70\% compression ratio and 76.08\% accuracy (accuracy drop is 0.10\% w.r.t to that of baseline) for DCASE-17 dataset. For DCASE-18 dataset, the compression ratio is 97\%, and accuracy is 63.78\% (accuracy gain is 1.27\% w.r.t that of baseline). The accuracy obtained usijg the proposed method is better than that obtained at different PCA-based explained variance ratio. This shows that the proposed method selects an appropriate compression ratio at a relatively marginal accuracy drop for DCASE-17 and improves performance for DCASE-18 compared to that obtained without any compression.

\noindent \textbf{Performance comparison:}
The performance comparison of the proposed framework with similar existing methods  that use raw-audio  or time-frequency representation (log-mel spectrogram or MFCCs) as input is shown in Figure \ref{fig: acc_comparison}. For DCASE-17, the comparison methods uses features such as spectro-temporal features \cite{Maka2017}, log-mel energies \cite{Heittola2017,Kim2017}. For DCASE-18, the comparison method includes log-mel energies \cite{Heittola2018}, MFCCs \cite{Purohit2018} and raw-waveform \cite{Purohit2018} as input to the classifier. The proposed method performs better than that of time-frequency representations and equivalent to that of existing raw-audio based framework.

\section{Conclusion}
\label{sec: conclusion}

This paper proposes a raw-audio based acoustic scene classification framework by incorporating different intermediate layers of SoundNet and reducing the dimensionality of each layer by identifying optimal dimensions using automatic compact dictionary learning framework.  The proposed framework performs better than that of existing time-frequency representation frameworks with an advantage of using raw-audio directly as input. Building an end-to-end network that uses learned pooling to represent intermediate layer features and reducing dimensionality using non-linear dimensionality reduction methods are some of the future directions.

%\end{document}
\bibliographystyle{IEEEtran}
\bibliography{refs.bib}
\end{sloppy}
\end{document}